\documentclass{revtex4}
\usepackage{amsfonts}
\usepackage{graphics}

\newcommand{\const}{\mathrm{const}}
\newcommand{\df}[2]{\frac{\partial #1}{\partial #2}}

\newcommand{\fig}[1]{fig.~\ref{#1}}
\newcommand{\figs}[1]{figs.~\ref{#1}}
\newcommand{\Fig}[1]{Figure~\ref{#1}}

\newcommand{\myfigure}[3]{\begin{figure}[p]
\centerline{\includegraphics{#1}}
\caption[]{#2}
\label{#3}
\end{figure}}

\begin{document}

\title{ Pursuit-evasion predator-prey waves in two spatial dimensions }

\author{V. N. Biktashev}
\affiliation{Department of Mathematical Sciences, University of Liverpool, Liverpool L69 7ZL, UK}
\author{J. Brindley}
\affiliation{Department of Applied Mathematics, University of Leeds, Leeds LS2 9JT, UK}
\author{A. V. Holden}
\affiliation{School of Biomedical Sciences, University of Leeds, Leeds LS2 9JT, UK}
\author{M. A. Tsyganov}
\affiliation{Institute of Theoretical and Experimental Biophysics, Pushchino, Moscow Region 142290, Russia}
\date{\today}

\begin{abstract}
We consider a spatially distributed population dynamics model with
excitable predator-prey dynamics, where species propagate in space due
to their taxis with respect to each other's gradient in addition to,
or instead of, their diffusive spread. Earlier, we have described new
phenomena in this model in one spatial dimension, not found in
analogous systems without taxis: reflecting and self-splitting
waves. Here we identify new phenomena in two spatial
dimensions: unusual patterns of meander of spirals, partial reflection
of waves, swelling wavetips, attachment of free wave ends to wave
backs, and as a result, a novel mechanism of self-supporting
complicated spatio-temporal activity, unknown in reaction-diffusion
population models.
\end{abstract}
\maketitle

\section{Introduction} 

% Leading paragraph
\textbf{
  To describe spatio-temporal dynamics of populations one needs, apart
  from local interaction of populations, to invoke some mechanism of
  spatial interaction, e.g spread of individuals in space. Often this
  is done in terms of diffusion, which is a macroscopic way of
  describing random, non-directed movement of individuals. This makes
  a population dynamics model a reaction-diffusion system, and pattern
  formation in reaction-diffusion systems is well researched. However,
  living organisms do not necessarily move randomly, and often show
  some directed movement in response to exogeneous factors, a
  behaviour usually described as taxis. Macroscopic representation of
  taxis yields equations different from reaction-diffusion systems,
  and pattern formation in such models has been studied much less. Our
  aim is to elucidate some new pattern formation mechanisms specific
  to this type of model.
}
 
In our previous work \cite{qs1,qs2}, we have described new phenomenon
observed numerically in a reaction-diffusion-taxis model, with
excitable predator-prey local kinetics \cite{TrBrind1}. In addition to
or instead of diffusion, we introduced taxis of each species on the
other's gradient: predators pursuing prey and prey escaping the
predators. We chose excitable rather than more traditional
limit-cycle predator-prey dynamics, as it is methodologically easier to
deal with solitary population waves than with wavetrains, and
identified some new features that are typical in our excitable
pursuit-evasion model, but unknown or very rare in reaction-diffusion
models of similar systems. These include ability of propagating waves to
penetrate through each other or reflect from impermeable boundaries
(rather than simply annihilate, as is typical for reaction-diffusion
waves), split and emit backward waves.

This was done in models with one spatial dimension. In most ecological
applications, however, two spatial dimensions are be more realistic. In
reaction-diffusion systems it is well known that two-dimensional
models can demonstrate a much wider variety of nontrivial solutions
than one-dimensional. Thus, in the present paper, we describe
phenomenology observed in two-dimensional simulations of
our pursuit-evasion excitable predator-prey model. As far as we are
aware, this is the first study of this sort, so rather than giving an 
exhaustive description, we have identified phenomena that
are qualitatively different from what is observed in two-dimensional
reaction-diffusion excitable systems. 

\section{The model}

We consider a two-dimensional version of the model, studied earlier in \cite{qs1,qs2}:
\begin{eqnarray}
\df{P}{t} &=& f(P,Z) + D\nabla^2{P} + h_-\nabla\left(P\nabla{Z}\right) , \nonumber\\
\df{Z}{t} &=& g(P,Z) + D\nabla^2{Z} - h_+\nabla\left(Z\nabla{P}\right) , \label{RDT}
\end{eqnarray}
where $P(x,y,t)$ and $Z(x,y,t)$ are biomass densities of the prey and
predator populations, $h_-$ is the coefficient determining the taxis
of prey down the gradient of predators (evasion), $h_+$ is the
coefficient determining the taxis of predators up the gradient of prey
(pursuit), $D$ is their diffusion coefficient, assumed equal for
simplicity, and $f$ and $g$ are ``kinetic'' terms, describing local
interaction of the species with each other. We assume that this local
interaction takes the Truscott-Brindley \cite{TrBrind1} form,
\begin{eqnarray}
f(P,Z) &=& \beta P(1-P) - Z P^2/(P^2+\nu^2), \nonumber \\ 
g(P,Z) &=& \gamma Z P^2/(P^2+\nu^2)-w Z ,      \label{TB} 
\end{eqnarray}
which describes excitable kinetics. Equations (\ref{RDT},\ref{TB}) are
solved numerically with the following parameter values: $\nu=0.07$,
$\beta=1$, $w=0.004$.  Parameters $\gamma$ and $D$ have been used in
two fixed combinations, (A) $\gamma=0.01$, $D=0.04$, and (B)
$\gamma=0.016$, $D=0$.  Parameters $h_-$, $h_+$ varied between
simulations.  From one-dimensional simulations, described in
\cite{qs2}, we know that in (A), purely reaction diffusion waves
without taxis terms ($h_-=h_+=0$) are possible, where as in (B), such
waves are not possible.

All two dimensional simulations were performed in a rectangular domain $(x,y)\in[0,L_x]\times[0,L_y]$, with no-flux boundary conditions,
\[
  \left.\df{(P,Z)}{x}\right|_{x=0,L_x}=\left.\df{(P,Z)}{y}\right|_{y=0,L_y}=0.
\]
In figures representing the results of the simulations, we designate
the domain size as $L_x\times L_y$, choice of $\gamma,D$ combination
as (A) or (B), the values of taxis coefficients as $(h_-,h_+)$, and
where the panels are presented with a regular time interval, it is
given as $\Delta t$.

\section{Pursuit-evasion waves in one dimension}

\subsection{Mechanism of propagation}

\myfigure{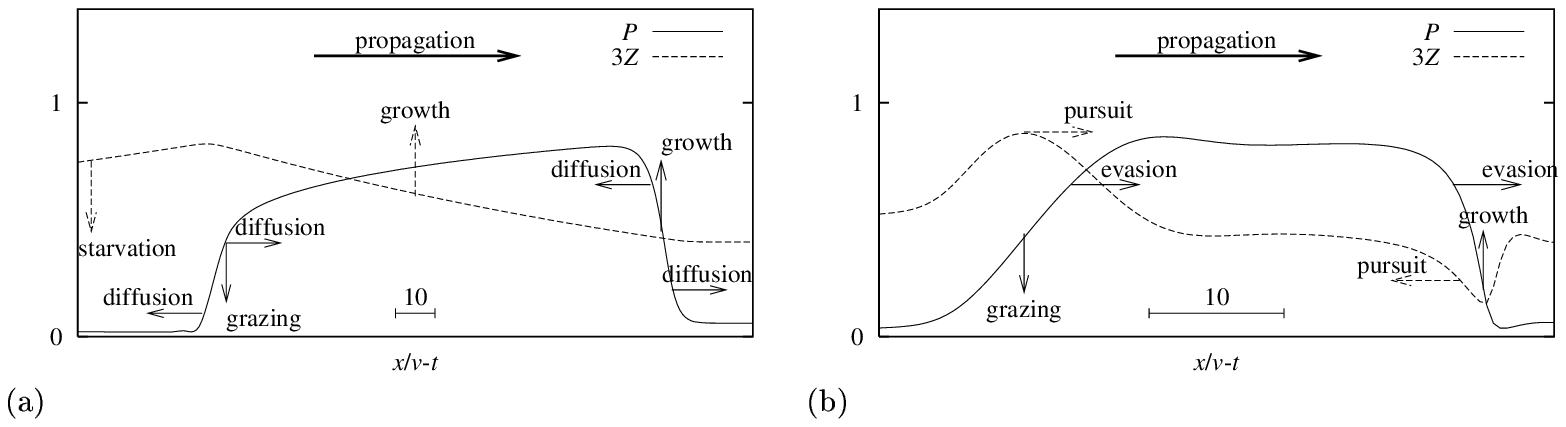}{
  A stationary propagating wave in 
  (a) reaction diffusion system ($D=0.04$, $h_{\pm}=0$), 
  (b) reaction-taxis system ($D=0$, $h_-=2$, $h_+=1$), 
  and $\gamma=0.01$ and other parameters standard in both cases.
  Arrows illustrate the main factors affecting the dynamics of the front and the back of the wave. 
  These include: prey dynamics (growth and decrease due to grazing pressure) in both (a) and (b), 
  diffusion of prey in (a), and 
  taxis (pursuit by predators and evasion by prey) in (b). 
}{profile}

It is useful to first recapitulate the one-dimensional results, and
we start with propagation of solitary waves.
\Fig{profile} shows typical profiles of steadily propagating waves in
a reaction-diffusion system and in a reaction-taxis system, and
schematically illustrates various mechanism involved. In the
reaction-diffusion wave (panel a), the density of predators changes
only slightly through the front. The propagation of the wave front is
mostly due to interaction between diffusion and nonlinear local
dynamics of the prey: influx of prey due to the diffusion triggers the
prey-escape mechanism, when the prey multiply faster than the
predators and therefore grow unchecked until reaching the carrying
capacity of the habitat. Between the front and the back of the wave
the prey and predators are in quasi-equilibrium, the predators slowly
consuming the prey and growing themselves. On the back, the
concentration of the predators is so large that the high concentration
of prey is no longer sustainable, and their dynamics are such that the
transition to their low state is sudden, and is therefore also
influenced by their diffusion, as at the front. The sequence of events
in the reaction-taxis wave is entirely different (panel b). The
advancing front of prey attracts the predators in the retrograde
direction. This leaves space relatively free of predators, which
triggers the prey-escape mechanism. Another mechanism, driving the
prey forward, is evasion, i.e. their taxis away from the predators.
This feedback loop of prey evading predators and predators pursuing
prey leads to spatial oscillations of both (similar to the way the
feedback loop of prey benefiting predators and predators harming prey
tends to produce their temporal oscillations).  The similarity with
reaction-diffusion waves is that the front peaks when the prey reach
their high quasi-equilibrium. The difference, apart from oscillatory
character of the taxis front, is that the density of predators does
not remain nearly constant during the front, but instead shows a
marked tip followed by a rise, due to predators first moving back
towards advancing prey, and then stopping that movement when the
gradient of prey is over. The events on the waveback are also quite
different. There we observe an advancing wave of predators pursuing
prey, and the receding wave of prey escaping predators. This recession
of prey is enhanced also by the fact that they are consumed by pursuing
predators. Thus the position of the back is not directly related to
the time taken by predators to grow to a certain concentration or to
consume a certain numbers of prey, but rather, by time and space
needed for the pursuit-evasion dynamics to form the back structure.
As a result, the duration of the taxis wave is much shorter
than that of the diffusion wave.

\subsection{Mechanism of reflection}

\myfigure{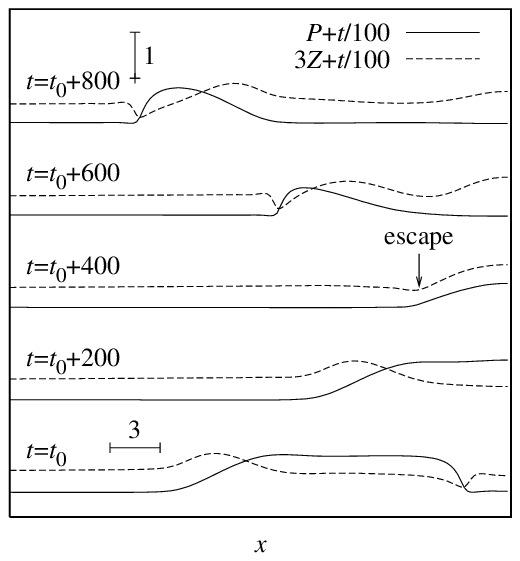}{
  Reflection of a taxis wave from an impermeable boundary ($D=0$, $h_-=2$, $h_+=1$, $\gamma=0.01$).
}{reflection}

\Fig{reflection} illustrates the process of reflection of the taxis
wave from an impermeable boundary. When the front of the wave arrives
at the boundary, the system there switches to the high-prey state,
close to the carrying capacity of the habitat, i.e. the maximal stationary
density of prey if predators are absent, which in our model is $P=1$
($t=t_0+200$).  This is similar to what happens in a
reaction-diffusion wave, and in both cases is mainly due to the
nonlinear dynamics of the prey population.  The difference comes when
the back of the excitation wave approaches the boundary
($t=t_0+200\dots t_0+400$), because in the taxis wave, unlike the
diffusion wave, the back of the wave profile is formed by prey evading
the peak of predators following them, rather than by consumption of prey by
predators.  As a result, in the taxis wave, the density of predators 
at the back of the wave is insufficient to reduce prey
density to a low level during the time that the tail actually
lasts.  Thus, when the evading prey are stopped by the impermeable
boundary, the predators catching up with them do not consume them all.
Instead, a substantial density of prey survives ($t=t_0+400$).
Those prey that survive after the maximum of the predator density has
passed them escape in the retrograde direction ($t=t_0+600$), and if
their number is large enough, initiate a new backward propagating wave
($t=t_0+800$).

Thus we see that the mechanism of reflection of taxis waves is
entirely different from known mechanisms of reflection of
reaction-diffusion waves, e.g. positive overlap of oscillatory tails
of the colliding waves discussed in \cite{Mornev96}, which required a
very close proximity of the local dynamics to the Hopf bifurcation. In
that case the reflection property was restricted to a narrow region in
the parameter space. With the present mechanism the reflection is
rather robust, both in terms of local dynamics, and of taxis and
diffusion coefficients \cite{qs2}.

\section{Spiral waves}

Amongst the most ubiquitous, dramatic and intensively studied
phenomena observed in 2D reaction-diffusion excitable systems are
re-entrant excitation waves, commonly known as spiral waves
\cite{Winfree-1972}. They are observed in a variety of experimental
systems, and in an even greater variety of mathematical models, 
that includes spatially extended mathematical models of predator-prey
interaction of the reaction-diffusion type
\cite{Sherratt-etal-2002}. Their significance stems from the fact that
they can occur in an event of a wave break in a propagating
excitation for whatever reason, and thus serve as a specific route 
spatio-temporal chaos. Thus, we have
started to study details of reaction-diffusion-taxis waves in 2D
through the effect that the taxis terms have on the spiral waves.

\subsection{Very small taxis can destabilize spirals}

\myfigure{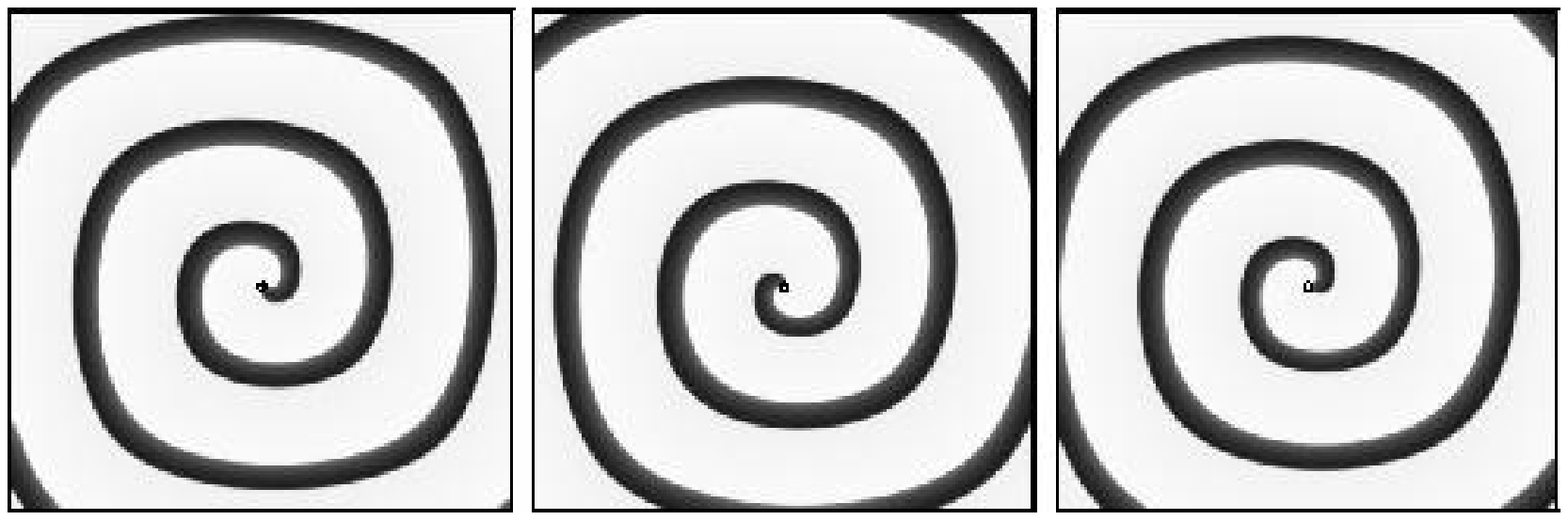}{
  Spiral wave in a purely diffusive medium, (A) $200\times200$ $(0,0)$, $\Delta t=150$.
  Here and on the next figures with spiral waves, the trajectory of the tip is superimposed.
  The tip is defined as a point where $P=0.49$ and $Z=0.21$; normally, there is only one such point per tip of a spiral wave.
}{spr-d}

\Fig{spr-d} shows a typical rigidly rotating spiral wave solution,
observed in the purely reaction-diffusion model. From the behaviour of
1D solutions\cite{qs2,qs2}, qualitatively new behaviour in 2D based 
on 1D phenomenology
(e.g. reflection from the boundaries) is expected when the taxis terms
are large enough in comparison with coefficient $D$ ($=0.04$ in our
simulations). The relevant values are of the order of $1$, which
corresponds to comparable fluxes generated by the diffusion and the
taxis terms, at the values of the variables at around the stable
equilibrium. Comparison of fluxes gives a reasonable way to compare
the relative significance of taxis and diffusion terms, as $h_{\pm}$
are coefficients of nonlinear terms and have different dimensionality
from that of $D$.

\myfigure{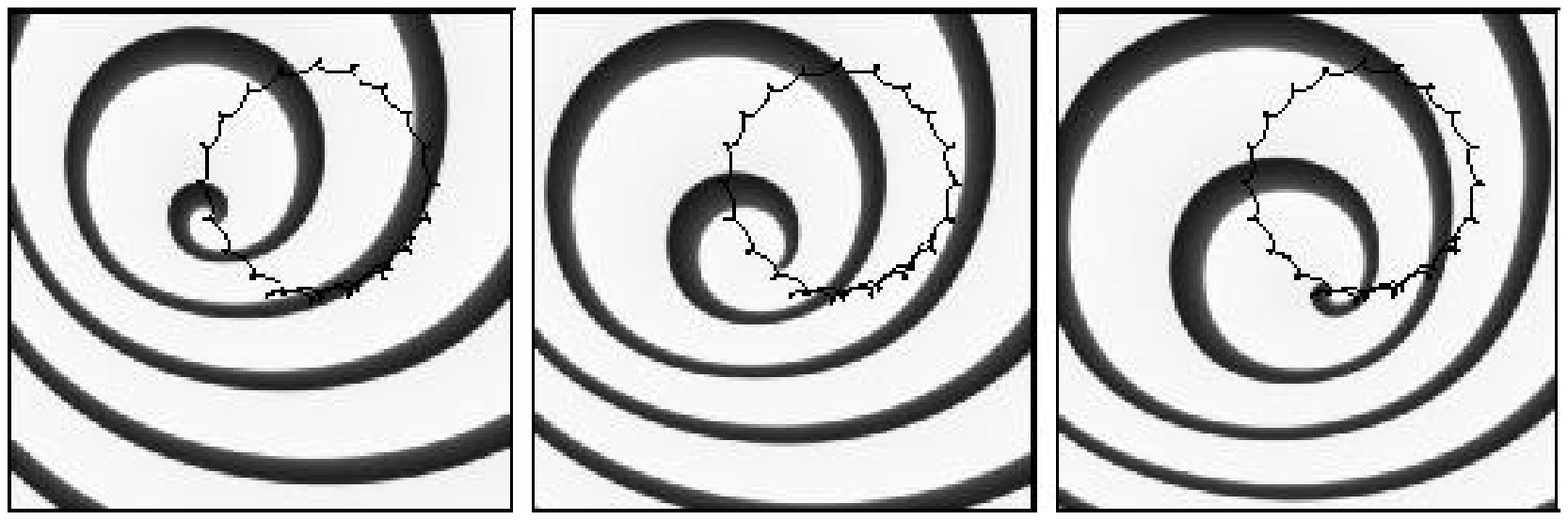}{
  Meandering spiral in a medium with small taxis coefficients,  
  (A) $200\times 200$ (0.05,0.05), $\Delta t=400$.
}{spr-dh05}

\Fig{spr-dh05} shows the behaviour of the reaction-diffusion-taxis
spiral wave at values of the taxis coefficients much smaller than
those at which reflection happens. Already these small taxis terms
change the behaviour of the spiral very significantly, as, with the
same local kinetics as the stationary spiral of \fig{spr-d}, it now
meanders wildly. This kind of non-stationary behaviour of spiral waves
is not unknown in excitable media, where the transition from
steady to meandering rotation is usually associated with change of the
parameters of the reaction; here we see such a transition for the same
reaction parameters but with the spatial terms altered.

\myfigure{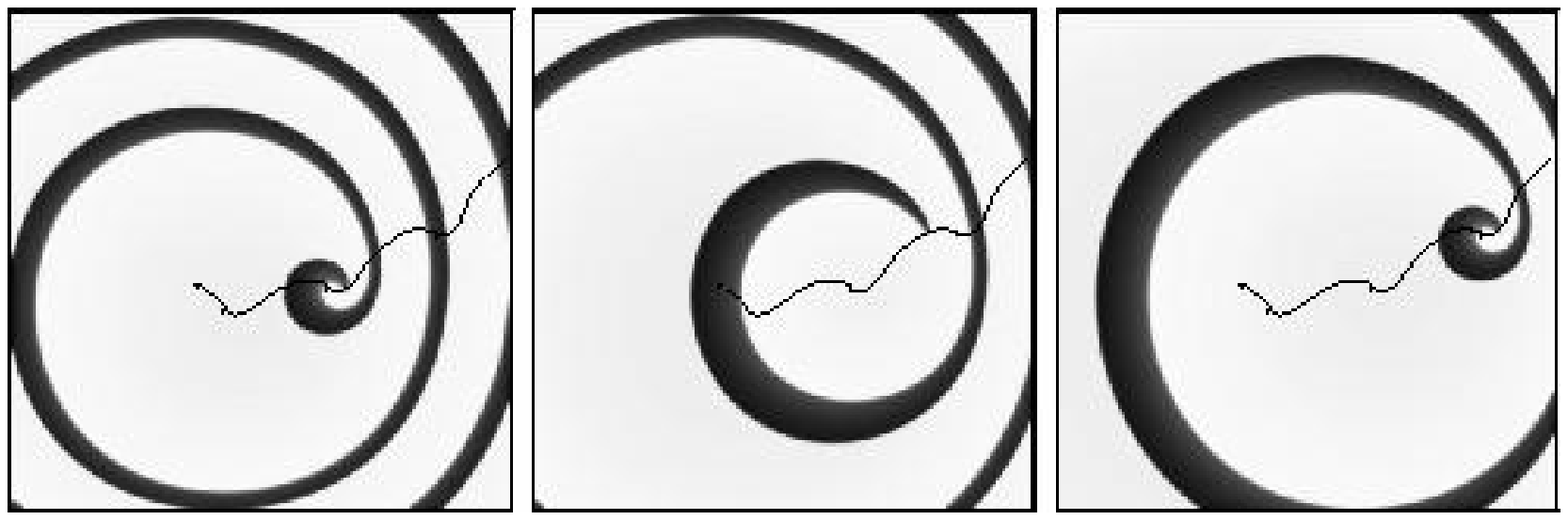}{
   Unstable spiral at larger taxis coefficients, (A) $200\times 200$
  (0.1,0.1), $\Delta t=225$.  
}{spr-dh01}

As the taxis coefficients increase, their effect on the spiral wave
dynamics increases further. \Fig{spr-dh01} shows that the spiral wave
meander becomes so extended that the spiral soon drifts out of the
medium and annihilates at the boundary. This happens at taxis
coefficients that are still much smaller than those needed for
quasi-soliton behaviour.

\subsection{Peculiar behaviour of the tip: waltzing spiral}

\myfigure{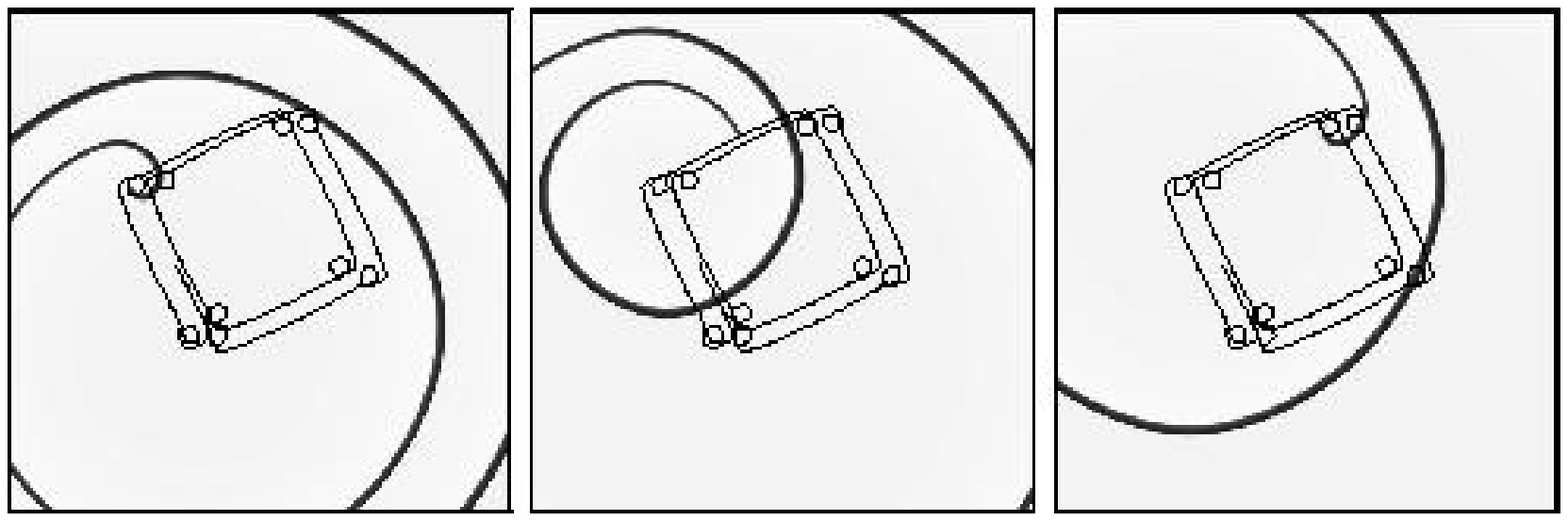}{
  Waltzing spiral wave, 
  (B) $250\times250$ (5,0), $\Delta t=450$.
}{spr-h07}

As the 1D studies have shown, if taxis terms are present, waves can
propagate without any diffusion. \Fig{spr-h07} shows a strikingly
different behaviour of a spiral wave in such a system in
2D. Apparently, the tip of the spiral periodically switches between
two meta-stable states, in one of which it curls up as in a typical
spiral wave in a strongly excitable reaction-diffusion system, and in
the other it propagates along a straight line orthogonally to the
front of the wave, as happens in an excitable system on the verge of
excitability. This results in a ``slow waltz'' trajectory, when the
spiral makes 5/4 of a turn, then moves along a straight line, then
makes another 5/4 of a turn, and so on. In total, the spiral makes
exactly five turns per each round of, and in the same direction as the
meander trajectory.

The ratio $1:5$ is not unique, as spiral waves in reaction diffusion
systems can demonstrate a continuous spectrum of these ratios as
kinetic parameters change \cite{Winfree-1991,Barkley-Kevrekidis-1994}.
It  fits well with theoretical considerations based on symmetry
considerations \cite{Barkley-1994,Biktashev-etal-1996,Wulff-1996,
Nicol-etal-2001,Golubitsky-etal-1997}, except that an exact four-fold
symmetry should, from the viewpoint of that theory, be
non-typical. Apparently in our case its appearance should be
attributed to a resonance with the slight anisotropy of the numerical
discretization.  The mechanism of the meander appears more exotic, in
which the tip trajectory consists of alternating pieces of very small and
large curvature, where the high-curvature pieces form more than one
full turn of the spiral wave. As far as we are aware this kind of spiral
wave meander has not been seen in reaction-diffusion systems.

% \section{When spiral waves do not start} 
% \section{Other characteristic phenomena}
\section{Alternative behaviour of the wave tips}

\myfigure{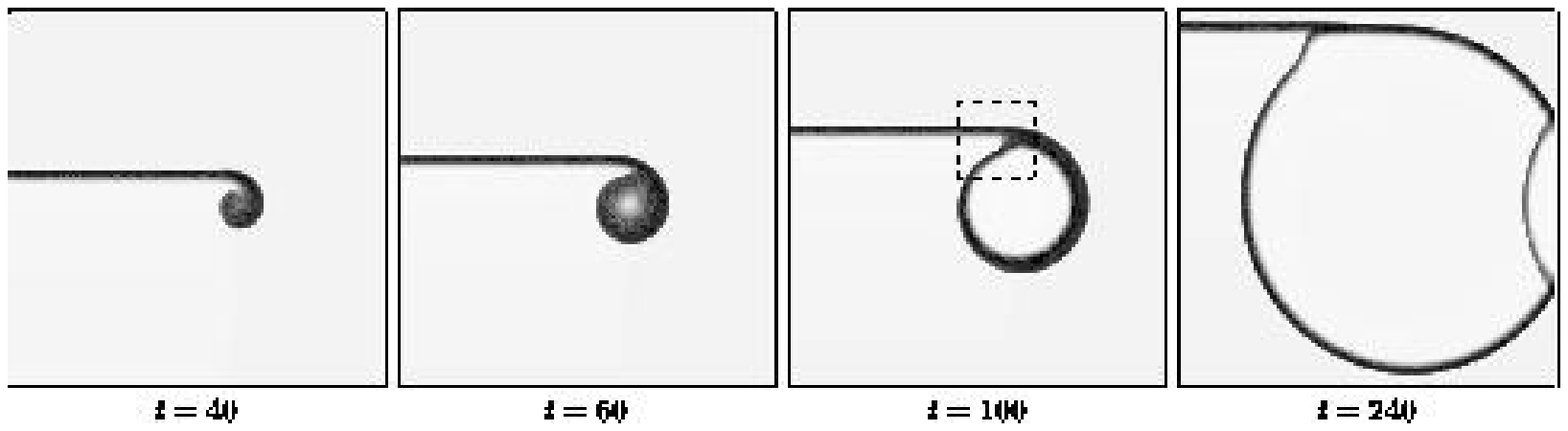}{
  Wavebreak that does not lead to spiral wave initiation, 
  (B) $250\times250$ (5,1).
  At $t=60$: predators have multiplied in the center of the circular prey patch, ending the ``swollen tip'' phase.
  At $t=100$: dashed rectangle is shown in detail below on \fig{touch}. 
  At $t=240$: a piece of the wave has reflected from the eastern boundary.
}{sp15-nonstart}

\myfigure{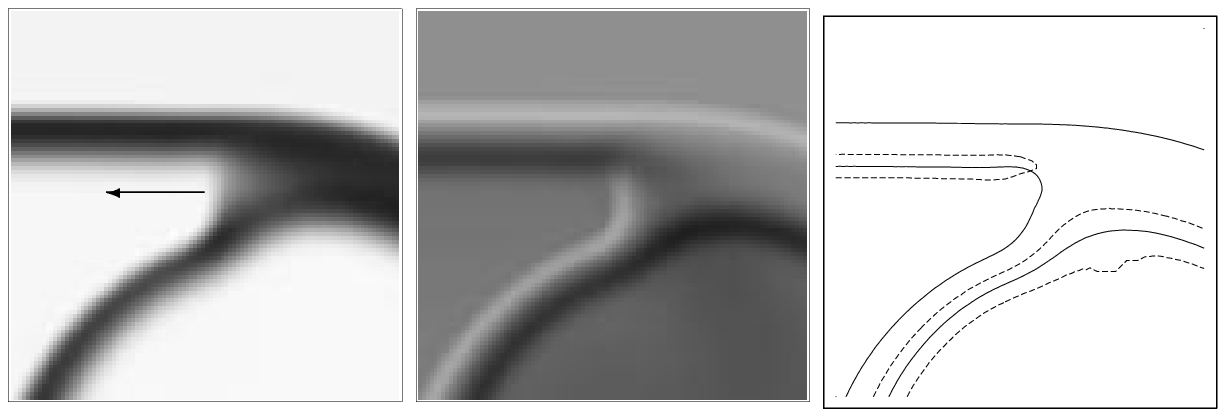}{
  Structure of the self-touching 
  (B) $250\times250$ (5,1).
 (a) distribution of $P$, (b) distribution of $Z$, 
 (c) isolines $P=\const$ (solid) and $Z=\const$ (dashed). 
}{touch}

In reaction-diffusion excitable systems, if an excitation wave is
broken, its tip will either grow (``germinate''), protrude and curl up
into a spiral wave, if the medium is ``strongly excitable'', or it may
shrink and retract, if it is ``weakly excitable''. \Fig{sp15-nonstart}
shows an example of a reaction-taxis excitable system, where neither
takes place. The tip of a broken wave protrudes, but does not curl up
into a spiral; instead new characteristic phenomena are observed. 

\subsection{Wavetip swelling}
See \fig{sp15-nonstart}, $t=40\dots 60$.  The consequence of the
wavebreak protruding may be seen as ``swelling'' of the tip, as it
goes through a ``blobbing'' stage, when an initially small oval-shape region of medium is
excited, which subsequently grows and reforms into circular expanding wave stage. A
phenomenological interpretation of this swelling phenomenon follows
from the empirical observation that it usually happens when the
evasion coefficient is large compared to the pursuit coefficient. Thus
the prey near the wavebreak have the capacity to escape from the
predators sideways. A subpopulation of prey then finds itself in a
region relatively free from predators. In this predator-free zone prey
start multiplying intensively and form a circular expanding region. As
there is no chasing wave of predators inside this circular region, the
high-prey state ends only when the predators multiply to large
densities, taking much longer than in a standard reaction-taxis
propagating wave where the end of the prey wave is determined by the
chasing wave of predators. So the swollen tip can grow to form a large
patch of prey, before turning into an annular shape wave, when the
back-structure of the taxis wave is formed. Subsequently, the predator
wave catches up with the prey and the circular expanding region
develops into a annulus-shaped expanding wave.

\subsection{Self-attachment}
See \fig{sp15-nonstart}, $t=60\dots 240$.  After the tip swelling
phase, no wavebreak can be identified based on the distribution of the
prey population. Nevertheless, a wavebreak, or tip, can be formally defined,
e.g. as an intersection of suitable isolines of the prey and predator
populations, as illustrated on \fig{touch}. For topological reasons,
such a tip cannot disappear, but has to move continuously until
reaching the medium boundary or annihilating with another tip of the
opposite chirality.  Such a ``tip'' is found near where the circular
wave, newly born from the swollen tip, touches the back of the
mother wave. Now the whole old+new wave can be seen as a single wave,
a free end of which is attached somewhere to its own back. This
could not happen in a reaction-diffusion medium, because of the refractory
period behind an excitation wave. In terms of the present population
dynamics system, the refractory period is characterised by such high
density of predators that triggering a prey-escape mechanism is not
feasible. The difference with taxis excitation waves is that they are
not characterised by significant difference in the predators' density
before and after the wave. To put it briefly, the prey wave finishes
not just when all prey are consumed by predators, but rather when
those that are not consumed have escaped forward. Similarly, the
predators' wave finishes not when all predators die out because of
starvation, but rather when all predators have moved forward in the pursuit
of prey. Thus the equilibrium concentration of predators and prey
behind the wave establishes rather quickly and due to a mechanism quite
different from reaction-diffusion waves. As a result, there is little
or no ``refractory tail'' behind a taxis wave, and this makes the
self-attachment of a wave's tip to its own back possible.

\myfigure{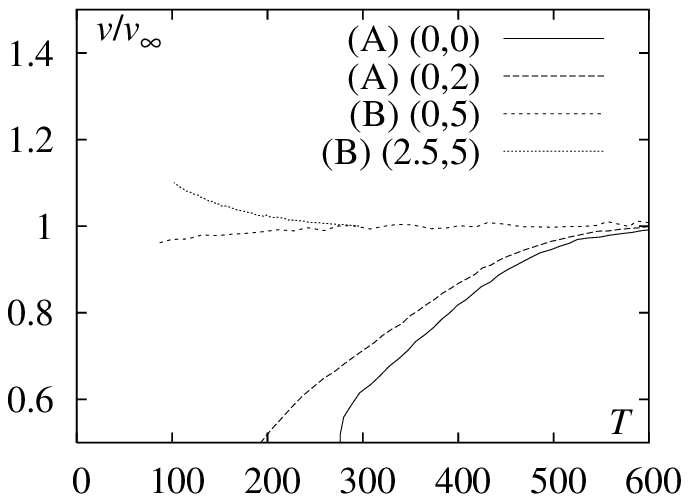}{
  Dispersion curves of periodic waves, as normalised velocity $v(L)/v(\infty)$ 
  vs time period $T=L/v(L)$, at different parameter values as indicated by the legend. 
  While reaction-diffusion waves go slower if frequent, taxis tends to speed frequent waves up. 
}{vl}

The structure of the attachment site is illustrated in more detail on
\fig{touch}. The high prey regions of the mother wave and the circular
wave are connected, whereas their high predator regions are disjoint,
or, more precisely, connected by an isthmus of much smaller
concentration of predators than elsewhere, because the predators at
the junction have been attracted backwards by the circular wave. Note
that the attached wave tip has to move considerably faster than a
plane solitary wave, as it participates simultaneously in the movement
of the mother wave, and the circular wave, which moves at an angle to
the mother wave. One more reason for such faster movement is 
the unusual nonlinear dispersion relationship for taxis waves, 
where speed increases as period decreses.
This property is illustrated on
\fig{vl}, which shows velocity of waves propagating on a
one-dimensional interval with periodic boundary conditions. Initially
a wave was started on an interval of a large length $L$. This length then
was decreased in small steps, after each of which sufficient time was allowed
for the circulating wave to approach its stationary velocity $v(L)$,
which was then recorded, and the next decrease of $L$ was made. To circumvent
the interpretation difficulty related to the difference of spatial
scales in diffusion and taxis terms, we plot not the absolute
velocity, but the velocity relative to the velocity of a solitary wave,
$v(\infty)$. For the same reason, the horizontal coordinate on
\fig{vl} is not the spatial period of waves $L$, but their temporal
period $T$ calculated as $L/v(L)$. 

Another way to view the mechanism of the attachment of the free end is through
kinetics of the prey, which multiply more efficiently because of the
gap in the predators' population at the site of the junction; 
the wave grows in the gap rather than
propagates into it (in fact, of course, prey growth is an essential
component of the wave propagation, so both explanations are valid).
 
\section{Partial reflection}

\myfigure{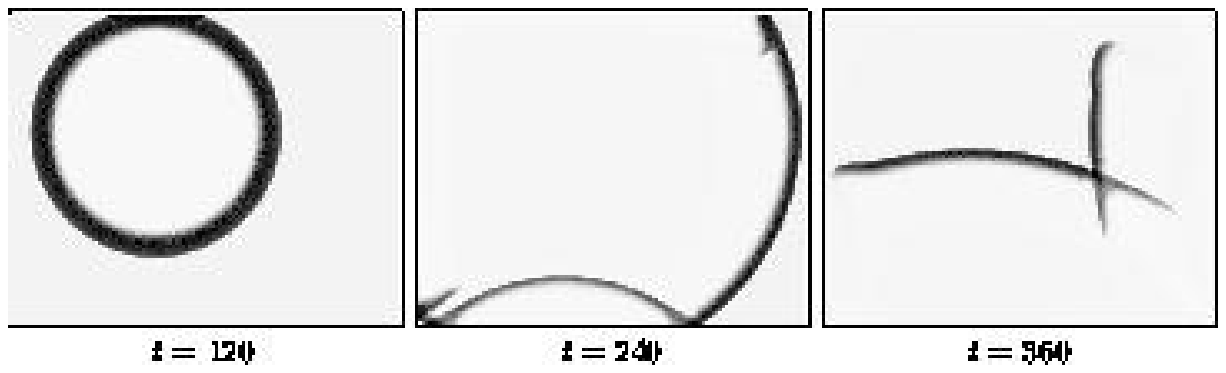}{
  Circular wave: some parts are reflected from the boundaries, some are not.
  (B) $200\times160$ (5,1).
}{rn016-begin}

As we have already mentioned, at appropriate taxis coefficients,
one-dimensional waves can reflect from the boundaries and penetrate
through each other. In two dimensions, there are new aspects
characterising impact of the waves with boundaries and each other, not
available in one dimension: the curvature of the waves at the moment
of impact, and angle of incidence. In reaction-diffusion systems, the
the wave propagation velocity is known to depend on the curvature of the wave.
We have found that in our reaction-taxis model, the curvature,
or angle of incidence, or both, affect the result of the impact,
i.e. whether the wave will be reflected from the medium boundary or
another wave, or will annihilate at it.  A representative example is
shown on \fig{rn016-begin}. A circular wave was initiated at an
asymmetrically located site within a rectangular region, so its
distances to all four boundaries were different. The result is that
the wave has mostly annihilated at the northern and western boundaries
(except small parts of the wave at the further ends of those
boundaries), which the wave reached first while having a higher
curvature. Then the wave has mostly reflected from the southern and
eastern boundaries, which it reached later while having a lower
curvature. The waves reflected from the eastern and southern
boundaries partly annihilate on impact, namely, where their collision
is more head-on. Meanwhile, where their collision is more slantwise,
the waves penetrate each other. One more factor that can affect the
outcome of the collision, is the ``age'' of the propagating wave. This
is related to delicate asymptotic properties of the taxis waves which
will be considered in more detail elsewhere\cite{qs5}. For the present
paper, the important observation is that a colliding wave may either
annihilate or penetrate/reflect in the same medium, depending on the
local
details of the collision. Therefore, the outcome of collision may be
different for different parts of the same colliding wave, which means
that the wave will be broken.

\section{Self-supporting activity}

We have identified two specifically two-dimensional phenomena that can
happen to pursuit-evasion waves: partial reflection/penetration, and
tip swelling. These two phenomena can work together to generate a
self-supporting spatio-temporal activity. The typical sequence of
events is: partially reflected/penetrated waves are broken, i.e. have
free ends. Unlike waves in reaction-diffusion systems, these free ends
do not curl up into spirals, but swell and produce circular waves with
the free end eventually attached to the back of its mother wave. 
These waves interact with each
other and with medium boundaries, where some of them are partially
penetrated/reflected, i.e. broken, which leads to new swollen tips,
and so this sequence of events repeats. We have, as a result, a
spatio-temporal ``chaos''. The mechanism described is different from
that known before in reaction-diffusion excitable systems, or
indeed in any other nonlinear spatially distributed systems. 
Examples of such self-supporting activities at different values of
parameters are shown on the series of figures,
\ref{rn016}--\ref{rc016}.
% \figs{rn016}{sp15}{spir25}{ring01}{rc016}
We present a broad collection of simulations, to show that this
mechanism is not restricted to a narrow region of parameters but is
rather typical. In \figs{rn016}, \ref{ring01} and \ref{rc016} initial
conditions were in the form of a single spot that produced initially a
circular wave, while in \figs{sp15} and \ref{spir25} initial
conditions were in the form of an artificially broken plane wave,
which in a reaction-diffusion system would generate a spiral
wave. \Fig{ring01} is for kinetics A, which supports waves,
particularly spiral waves in a reaction-diffusion system without
taxis, whereas the rest of this set are for kinetics B, with various
combinations of taxis coefficients, which does not support either
spiral or any other type of waves without taxis. Kinetics A of
\fig{ring01} shows strong waves, with relatively thick stripes of prey
and very large swollen tips, whereas kinetics B in \fig{rc016}
produces relatively thin waves with rather modest swollen tip (their
traces can be noticed e.g. at the $t=620$ panel). The self-supporting
activity can last long, but not necessarily for ever, as there is
always a possibility that all waves will annihilate on collisions so that
activity survives, as happens in \fig{rc016}. Obviously the
probability of that is higher in a smaller system. For instance, in
\fig{rc016}, the square symmetry of the problem means that the
effective size of the system is eight time smaller than the area of
the square.

%%%%%%%%%%%%%%%%%%%%%%%%%%%%%%%%%%%%%%%
\myfigure{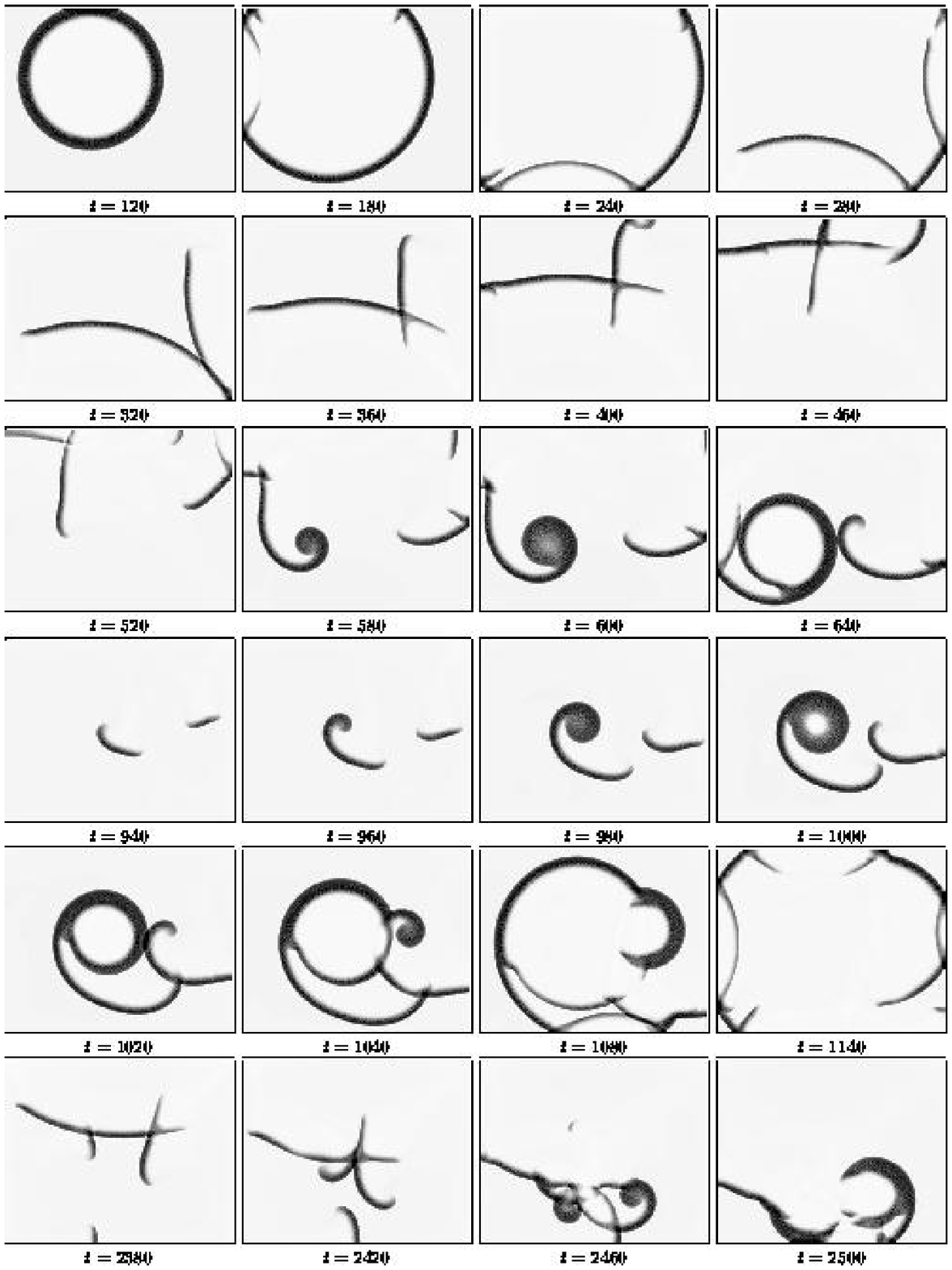}{
  Self-supporting activity. 
  Continuation of \fig{rn016-begin}, 
  (B) $200\times160$ (5,1).
}{rn016}

%%%%%%%%%%%%%%%%%%%%%%%%%%%%%%%%%%%%%%%

\myfigure{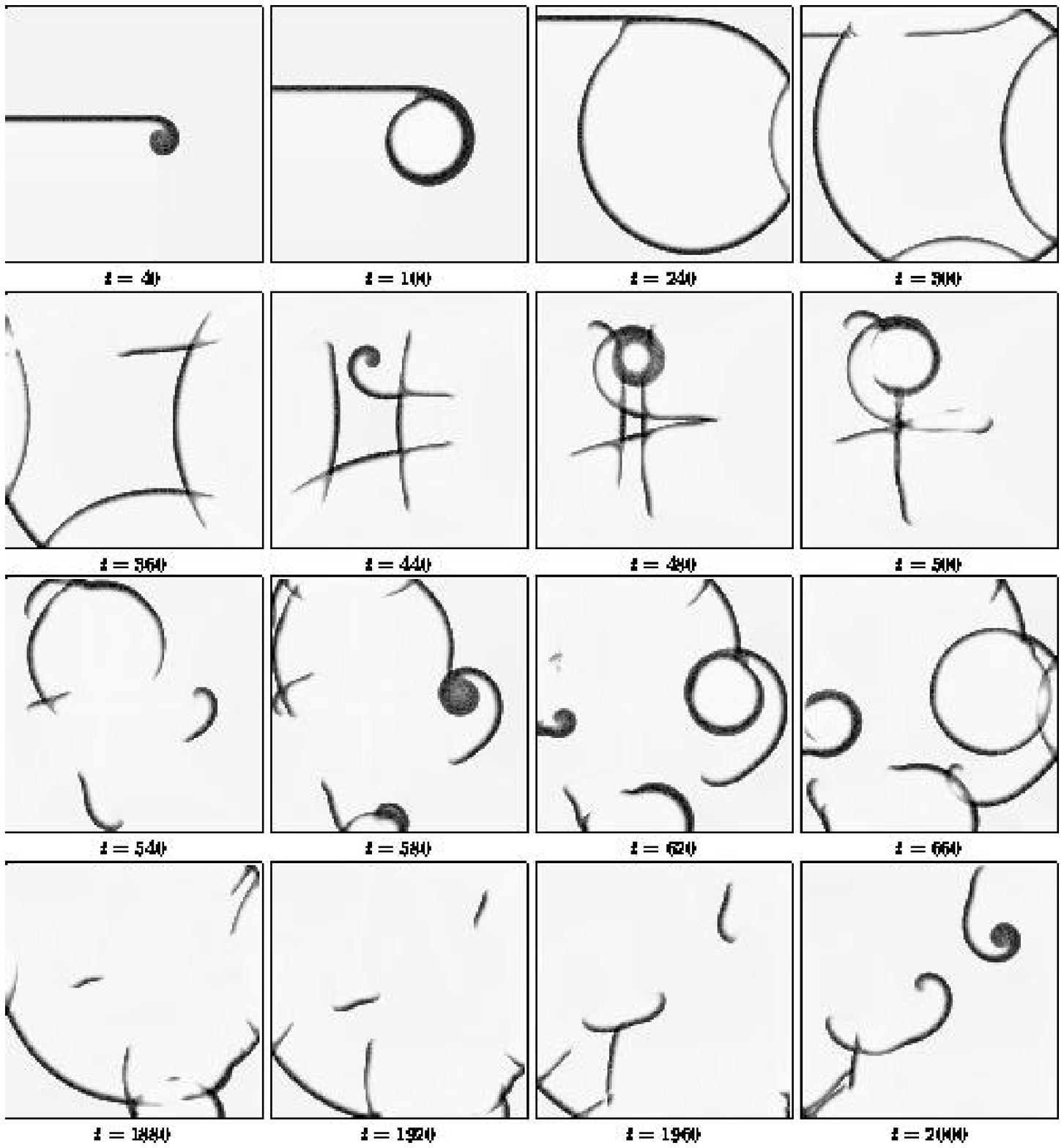}{
  Self-supporting activity. 
  Continuation of \fig{sp15-nonstart}, 
  (B) $250\times250$ $(5,1)$.
}{sp15}

\myfigure{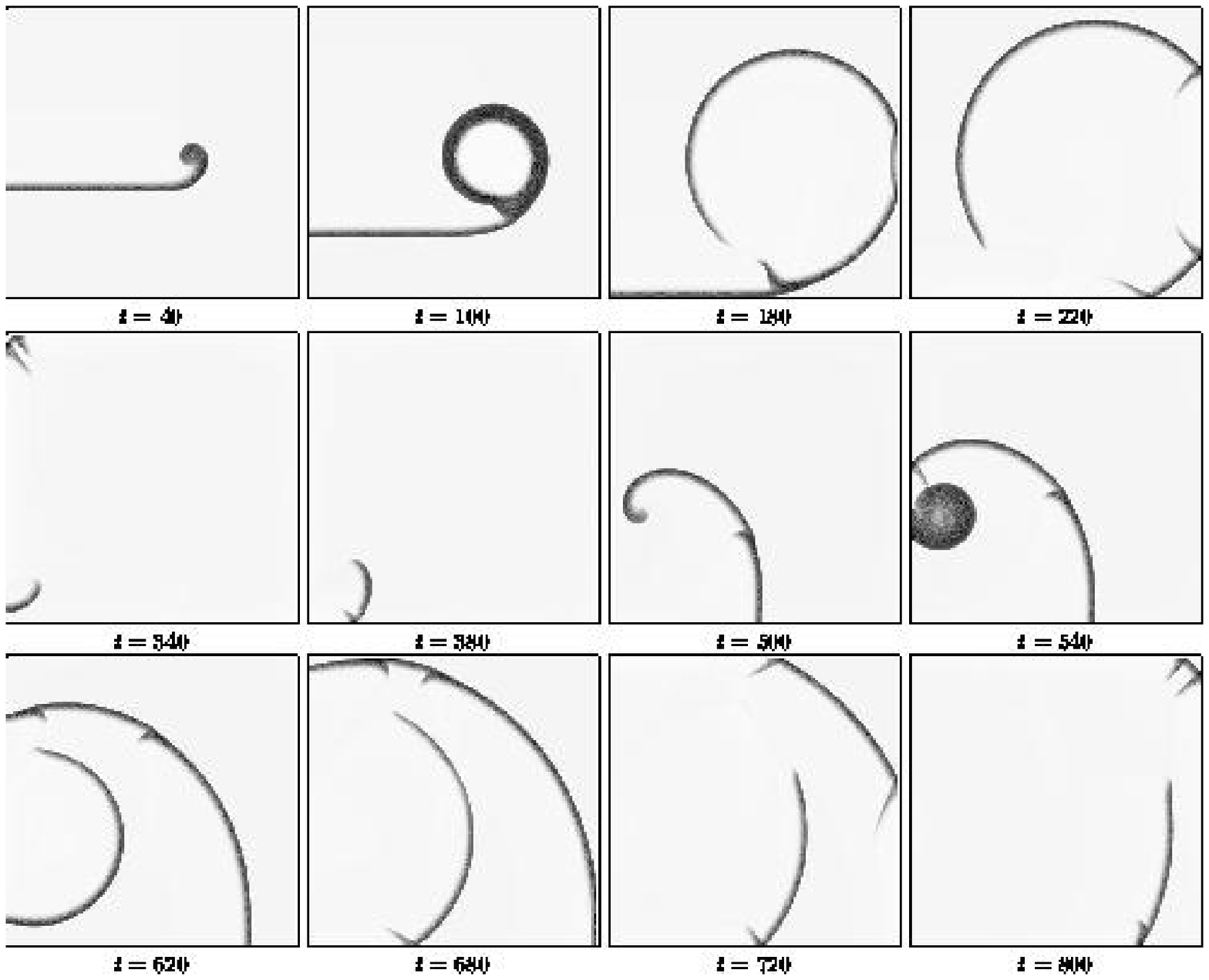}{
  Self-supporting activity. 
  (B) $250\times250$ $(5,2)$.
}{spir25}

\myfigure{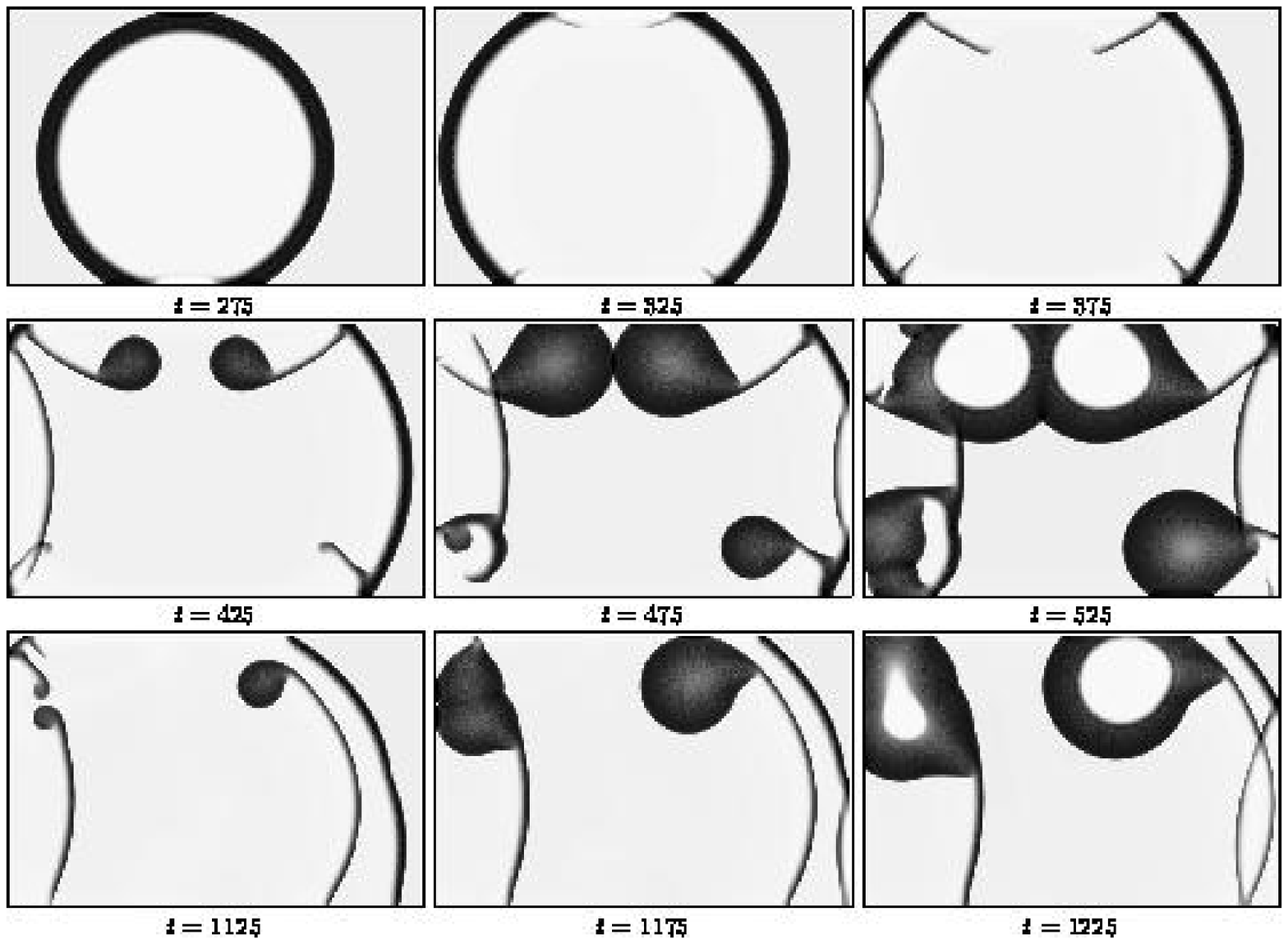}{
  Self supporting activity, 
  (A) $350\times230$ $(2,1.5)$.
}{ring01}

\myfigure{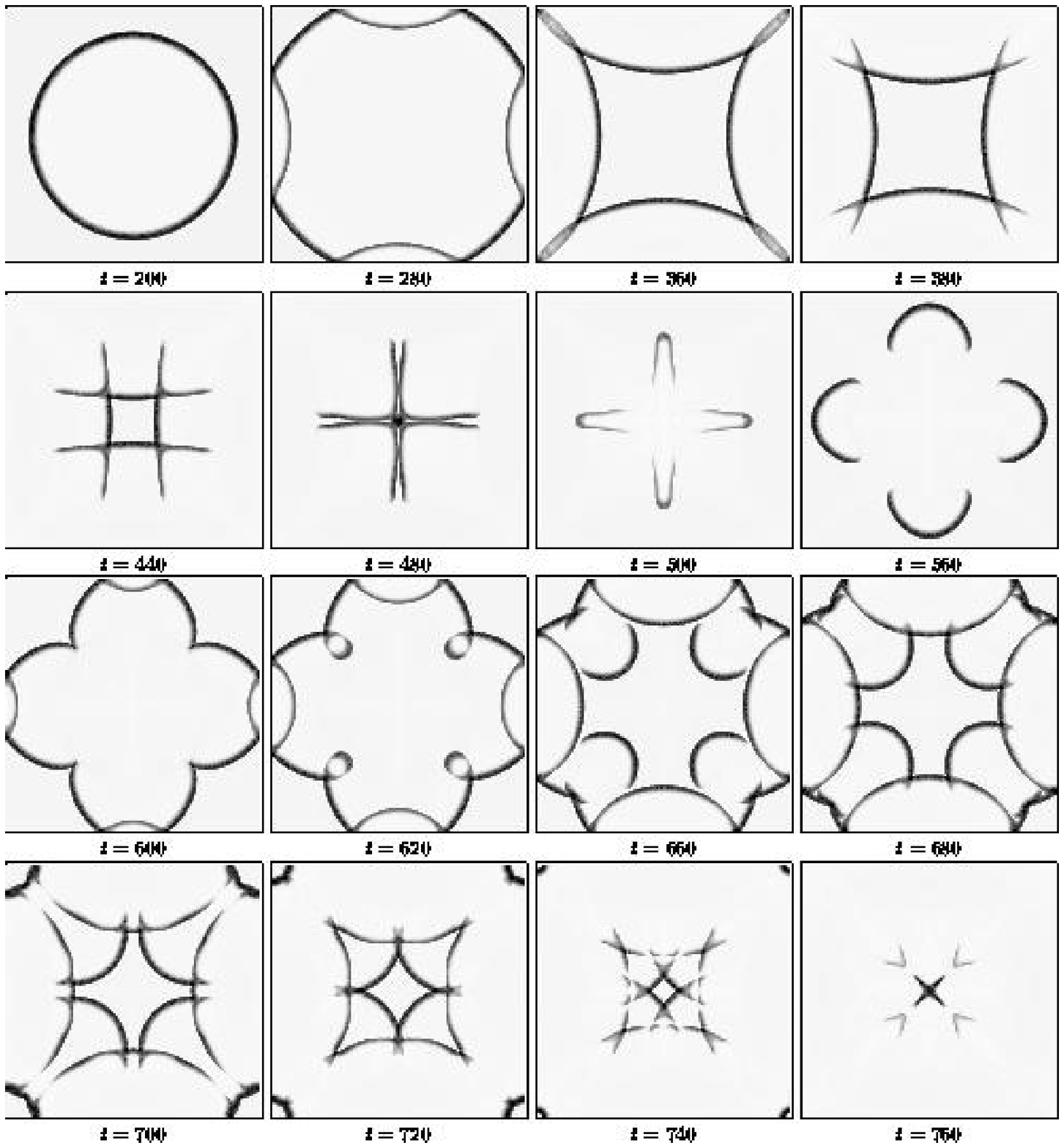}{
  Self supporting activity, in a symmetric setting. 
  This time the activity stops eventually, when all available waves coincidentally 
  annihilate at the same time. 
  (B) $250\times250$, $(5,1)$.
}{rc016}

\section{Conclusions}

We have demonstrated that two-dimensional excitable
reaction-diffusion-taxis systems can exhibit new properties,
unavailable in one-dimensional systems of the same kind, and different
from two-dimensional purely reaction-diffusion systems with the same
kinetics. Even relatively small taxis terms can make spiral waves very
unstable in conditions where they would be stable without such
terms. If the taxis (pursuit/evasion) terms are large enough, compared
to diffusion, to cause quasi-soliton behaviour in 1D, then the 2D
behaviour is entirely different and is no longer dominated by
processes of spiral waves generation and births, as is typical for
reaction-diffusion systems. Instead, a new type of self-supporting
activity takes place, which is mediated by three qualitatively new
types of events,
\begin{itemize}
\item Partial reflection of waves from boundaries, or their partial
penetration through each other, which produces broken waves, and
\item ``Swollen tips'', i.e. circular wave sources, produced by free
ends of broken waves.
\item Attachment of free ends of broken waves to the wavebacks. 
\end{itemize}
The first of these phenomena is due to sensitive dependence of
quasi-soliton behaviour on the circumstances of collision. The second
is due to the increased possibility of the sideward escape of prey
from the head of the propagating taxis wave near the wave break, when
their mobility is higher than that of the predators. The third is due
to short supply of predators at the site of attachment, which allows
prey to multiply quicker to seal the gap. All three phenomena are
related to the peculiar feature of the reaction-taxis waves, the
virtual absence of the refractory tail, which is a prominent feature
of reaction-diffusion excitation waves.

The value of this study is twofold. First, this illustrates the
importance of taking into account of directed movement of species in
modelling spatially distributed interacting populations, as this can
produce completely different phenomena from those that occur in
reaction-diffusion systems. Second, this is interesting in a broader
nonlinear science context, as an example of a new type of nonlinear
dissipative waves, with some properties similar to those known
before, and some completely new.

As we noted earlier, quasi-soliton behaviour can be observed in some
reaction-diffusion excitable systems if its kinetics are close to a
Hopf bifurcation\cite{Mornev96}. In two spatial dimensions, such
systems also demonstrate unusual types of behaviour, such as
converging, concave spirals
\cite{Mornev-etal-1997,Mornev-etal-2000,Mornev-etal-2003}, 
but again in very limited regions of parameter space.

\section*{Acknowledgments} This study was supported in part by EPSRC grant GR/S08664/01 (UK) and RFBR grant 03-01-00673 (Russia).

% \bibliographystyle{unsrt}
% \bibliography{qs3} 

\end{document}